\documentclass[%
 reprint,
 amsmath,amssymb,
 aps,
]{revtex4-1}

\usepackage{graphicx}
\usepackage{dcolumn}
\usepackage{bm}

\begin{document}

\preprint{APS/123-QED}

\title{Efficient and accurate analysis of photon density of states for two-dimensional photonic crystals  with omnidirectional light propagation}


\author{Ruei-Fu Jao$^{1}$}
\author{Ming-Chieh Lin$^{2,}$}
 \email{mclin@hanyang.ac.kr}
\altaffiliation[Also at ]{the Institute
for Pulsed Power and Microwave Technology, Karlsruhe Institute of Technology, Germany as a visiting scholar.}
\affiliation{%
$^{1}$School of Information Technology, Guangdong Industry Polytechnic, Guangzhou, Guangdong 510300, P. R. China\\
$^{2}$Multidisciplinary Computational Laboratory, Department of Electrical and Biomedical Engineering, Hanyang University, Seoul 04763, Korea\\
}%




\date{\today}

\begin{abstract}
Omnidirectional light propagation in two-dimensional (2D) photonic crystals (PCs) has been investigated by extending the formerly developed 2D finite element analysis (FEA) of in-plane light propagation in which the corresponding band structure (BS) and photon density of states (PDOS) of 2D PCs with complex geometry configurations had been calculated more accurately by using an adaptive FEA in real space for both the transverse electric (TE) and transverse magnetic (TM) modes. In this work, by adopting a waveguiding theory under the consideration of translational symmetry, the omnidirectional PDOS corresponding to both the radiative and evanescent waves can be calculated accurately and efficiently based on the in-plane dispersion relations of both TE and TM modes within the irreducible Brillouin zone. We demonstrate that the ¡°complete band gaps¡± shown by previous work considering only the radiative modes will be closed by including the contributions of the evanescent modes. These results are of general importance and relevance to the spontaneous emission by an atom or to dipole radiation in 2D periodic structures. In addition, it may serve as an efficient approach to identifying the existence of a complete photonic band gap in a 2D PC instead of using time-consuming 3D BS calculations.
\end{abstract}

\maketitle


\section{\label{sec:level1}INTRODUCTION}

In the past three decades, photonic crystals (PCs) have attracted much attention \cite{1,2}. Photonic crystals, according to the dimension of the periodicity, are divided into three categories, namely one-, two-, and three-dimensional (3D) crystals. Periodic dielectric materials are characterized by photonic band gaps (PBGs). A PBG can prohibit the propagation of electromagnetic (EM) waves whose frequencies fall within the band gap region. These materials are expected to have many applications in optoelectronics and optical communications. Controlling the optical properties of materials has become a key issue in material engineering. It was proposed that the emission of EM radiation can be modified by the environment \cite{3,4}. Several environments such as metallic cavities \cite{5}, dielectric cavities \cite{6}, and superlattices \cite{7,8,9,10,11,12} have been studied. The environmental effects have been described by the photon density of states (PDOS) which is related to the transition rate of the Fermi golden rule. In principle, a complete PBG along all dimensions in space can be best realized in a 3D system. However, the difficulty in fabricating such 3D crystals with PBGs in the optical regime prohibits the progression of many applications. Many studies in 2D PCs have been mainly focused on the in-plane propagation of EM waves \cite{13,14,15,16,17,18}. In our previous work \cite{13}, we analyzed the in-plane light propagation in 2D PCs and demonstrated that the corresponding band structure (BS) and PDOS of 2D PCs can be calculated more accurately by using an adaptive finite element analysis (FEA) in real space for both the transverse electric (TE) and transverse magnetic (TM) modes, with even more complex geometry configurations. Various types of period structures exhibit PBGs. However, for some applications, the investigation of an omnidirectional light propagation is crucial. Previous studies showed the possibility of having omnidirectional absolute band gaps in some 2D crystal structures by adopting the off-plane wave vector $k_{z} = k_{0}sin\theta$, where $k_{0} = \omega/c$ \cite{19,20}. Theoretically, there are no band gaps for propagation in the $z$ direction. As $k_{z}$ increases, the modes decouple and the bandwidth shrinks to zero \cite{21,22,23,24,25,26}.

In this work, by adopting a waveguiding theory under the consideration of translational symmetry and extending the in-plane model \cite{13}, the omnidirectional PDOS corresponding to both the radiative and evanescent waves can be calculated accurately and efficiently based on the in-plane dispersion relations within the irreducible Brillouin zone \cite{13}. In the following, we first provide the detailed formulations in our simulation model in which the contributions of the total PDOS from both the radiative and evanescent waves for different polarization characteristics including both the TE and TM modes can be distinguished, then the validation of our approach, and finally demonstrate that the ¡°complete band gaps¡± shown by previous work considering only the radiative modes will be closed by including the contributions of the evanescent modes.

\section{\label{sec:level1}FORMULATION}

The propagation of light in a photonic crystal can be studied by solving Maxwell's equations. Figure 1 shows the schematics of omnidirectional light propagation in 2D PCs for (a) a triangular lattice and (b) a square lattice with $k_{z} = k_{0}sin\theta$ (black solid line), and the in-plane light propagation with $k_{z} = 0$ (blue dashed line), where $\theta$ is the off-plane incident angle.  For time-harmonic fields, it is convenient to use phasor notation. Maxwell's equations lead to the wave equations, or the master equations:

\begin{equation}
\nabla \times \left[ \frac{1}{\mu(r)}\nabla \times \vec{E}(r) \right] - \omega^{2}\varepsilon(r)\vec{E}(r)=0
\end{equation}
and
\begin{equation}
\nabla \times \left[ \frac{1}{\varepsilon(r)}\nabla \times \vec{H}(r) \right] - \omega^{2}\mu(r)\vec{H}(r)=0,
\end{equation}

\begin{figure}
\includegraphics[width=0.5\textwidth]{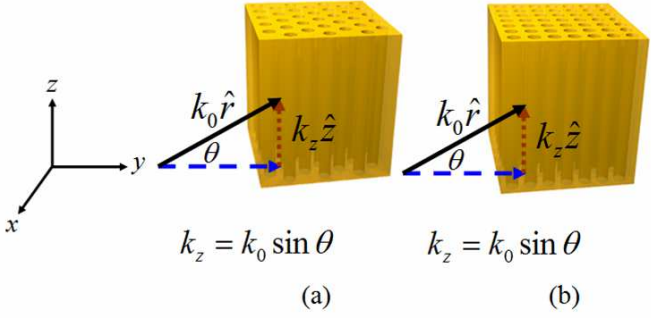}
\\[.3cm]
\caption{\label{fig:epsart} Schematics of omnidirectional light propagation in 2D PCs for (a) a triangular lattice and (b) a square lattice with $k_{z} = k_{0}sin\theta$ (black solid line) and the in-plane light propagation with $k_{z} = 0$ (blue dashed line), where $\theta$ is the off-plane incident angle.}
\end{figure}

where $\varepsilon(r)$ and $\mu(r)$ are the permittivity and permeability functions of the PCs, respectively, and $\omega$ is the angular eigen-frequency. In a 2D periodic system, the dielectric function is a periodic function of $x$ and $y$. We assume that the materials are linear, homogeneous, isotropic, lossless, and nonmagnetic. We have

\begin{equation}
\varepsilon_{r}(x,y) = \begin{cases} \varepsilon_{a}, & \text{$x,y$} \in $ air region$  \\\\
\varepsilon_{d} , & \text{$x,y$} \in $ dielectric region$ \end{cases}
\end{equation}

where $\varepsilon_{r}(x,y)$ is the dielectric function profile, and $\varepsilon_{a}$ and $\varepsilon_{d}$ are the dielectric constants of the air and dielectric regions, respectively. The two master equations are reduced to two homogeneous Helmholtz's equations for the air (dielectric) region:

\begin{equation}
\nabla^{2}\left\{
\begin{array}{cc}
\vec{E}(r) \\
\vec{H}(r)
\end{array} \right\}+\frac{\omega^{2}}{c^{2}}\varepsilon_{a(d)}\left\{
\begin{array}{cc}
\vec{E}(r) \\
\vec{H}(r)
\end{array} \right\}=0,
\end{equation}

A two-dimensional photonic crystal is periodic in two directions $(x,y)$ and homogeneous in the third one $z$.For light propagating in the system retaining translational symmetry, we can separate the modes into two independent polarizations, TM and TE modes, and consider the band structures and photon density of states accordingly. Based on the theory of waveguiding, the propagation properties of TM and TE modes can be characterized by the field components parallel to the rods or along the $z$ direction, $E_{z}(x,y)$ and $H_{z}(x,y)$, respectively. The corresponding Helmholtz's equations, the $z$ components of Eq. (4), for the air (dielectric) region can be rearranged as:

\begin{equation}
\left [
\frac{\partial^{2}}{\partial x^{2}}+\frac{\partial^{2}}{\partial y^{2}}+ \frac{\partial^{2}}{\partial z^{2}}
+\frac{\omega^{2}}{c^{2}}\varepsilon_{a(d)}
\right]\left\{
\begin{array}{cc}
\vec{E}(r) \\
\vec{H}(r)
\end{array} \right\}=0.
\end{equation}

As the system has translational symmetry along the $z$-axis, we can assume the longitudinal wave functions to be a plane wave, $exp(-ik_{z}z)$. By using separation of variables, Eq. (5) can be split into transverse and longitudinal parts and the problem can be simplified as solving Helmholtz's equations in the $xy$-plane. We obtain

\begin{equation}
\left [
\frac{\partial^{2}}{\partial x^{2}}+\frac{\partial^{2}}{\partial y^{2}}+
\left (
\frac{\omega^{2}}{c^{2}}\varepsilon_{a(d)}-k_{z}^{2}
\right)
\right]\left\{
\begin{array}{cc}
\vec{E}(r) \\
\vec{H}(r)
\end{array} \right\}=0.
\end{equation}

The in-plane propagation ($k_{z}=0$) can be considered as a limiting case or a cut-off condition of the omnidirectional propagation ($k_{z}\ge 0$), as shown in Fig. 1. Therefore, we can solve the following 2D Helmholtz's equations for the cut-off eigenvalues:

\begin{equation}
\left [
\frac{\partial^{2}}{\partial x^{2}}+\frac{\partial^{2}}{\partial y^{2}}+
\left (
\frac{\omega^{2}}{c^{2}}\varepsilon_{a(d)}
\right)
\right]\left\{
\begin{array}{cc}
\vec{E}(r) \\
\vec{H}(r)
\end{array} \right\}=0,
\end{equation}

where $\omega_{c}$ is the cut-off angular eigen-frequency for the omnidirectional propagating waves. Then the corresponding dispersion relations for the omnidirectional light propagation in the 2D PCs can be determined by

\begin{equation}
k_{z}^{2}=\frac{\omega^{2}-\omega_{c}^{2}}{c^{2}}\varepsilon_{r}.
\end{equation}

To perform the 3D PDOS calculations, we construct two equifrequency regions $\omega(k_{x},k_{y},k_{z})=\omega$ and $\omega(k_{x},k_{y},k_{z})=\omega+ d\omega$, where $\omega$ is an arbitrary value of the angular frequency and $d\omega$ is an infinitesimal increment \cite{12,13}. The differential volume element in $\textbf{k}$ space is $dV_{k}=dk_{x}dk_{y}dk_{z}$ Finally, according to the definition, the expression for the total PDOS is $dN(\omega)\equiv D(\omega)d\omega $:

\begin{equation}
D(\omega)=\frac{V\omega \sqrt {\mu_{r}\varepsilon_{r}}}{8\pi^{3}c}\int_{\omega_{k}}
\frac{\omega}{\sqrt {\omega^2-\omega_{c}^{2}}}dk_{x}dk_{y},
\end{equation}

where \emph{V} is the volume of the system in real space and $\mu_{r}=1$ for nonmagnetic material.

\section{\label{sec:level1}RESULTS AND DISCUSSION}

The omnidirectional PDOS can be obtained by employing Eq. (9) in which we perform a numerical integration of the 2D in-plane dispersion relations, Eq. (7). According to the waveguiding theory, the contributions to the total PDOS from both the TE and TM modes can be considered and calculated independently, so that the polarization characteristics of PDOS can be distinguished. In addition, using the formulation of the critical angle from Snell's law, the PDOS of the radiative $[cos^{-1} (\omega_{c}/\omega)\leq sin^{-1} (\varepsilon_{a}/\varepsilon_{d})^{1/2}]$ and evanescent $[cos^{-1} (\omega_{c}/\omega) > sin^{-1} (\varepsilon_{a}/\varepsilon_{d})^{1/2}]$ modes can be calculated separately while performing the numerical integration. One should note that the in-plane dispersion relations, Eq. (7), are calculated using the adaptive finite element method (FEM) in real space which had been demonstrated to be very accurate \cite{13}. Therefore, as the calculation of total 3D PDOS for a two-dimensional photonic crystal is based on the adaptive FEM and numerical integration, the accurate evaluation of the 3D PDOS in our extended model is justified.

In order to validate the extended model, we consider the omnidirectional light propagation in an inhomogeneous, linear, and nonmagnetic medium and employ a 2D PC model with a triangular lattice of air cylinders etched into silicon $(\varepsilon_{r}=11.90)$ at a filling ratio of $67\%$ air as calculated by the plane wave expansion method (PWEM) in Ref. \cite{21}, similar to the schematic shown in Fig. 1(a). Figure 2 shows the comparisons of 3D total PDOS of the 2D PC calculated using the FEM and PWEM, represented by the black solid line and the green open circles, respectively. As one can see, the 3D total PDOS calculated by our method is contributed from both the radiative (red dotted line) and evanescent (blue dashed line) modes and the results showing no complete 3D PBG are in good agreement with those calculated by the PWEM. The PWEM is based on the Bloch-Floquet theorem, which states that eigensolutions of differential equations with periodic coefficients may be expressed as a product of plane waves and lattice-periodic functions. Consequently, all periodic functions are expanded into appropriate Fourier series. Inserting these expansions into the differential equation results in an infinite matrix-eigenvalue problem, which, suitably truncated, provides the eigenfrequencies and expansion coefficients for the eigenfunctions. In the framework of PWEM, the 3D total PDOS calculation is based on fully 3D PBG computations and the Fourier coefficients are determined by integrating over the 3D Wigner-Seitz cell.  Therefore, there is no decoupling of the two transverse polarizations and the full 3D vector problem has to be solved \cite{21}. As a Fourier-based method, it is not only time-consuming but also suffers from several problems. For instance, the dielectric function is discontinuous, so Fourier-type expansions converge slowly. Furthermore, it was found that the discontinuous nature of the dielectric function severely limits the accuracy of the PWEM \cite{27}.

\begin{figure}
\includegraphics[width=0.47\textwidth]{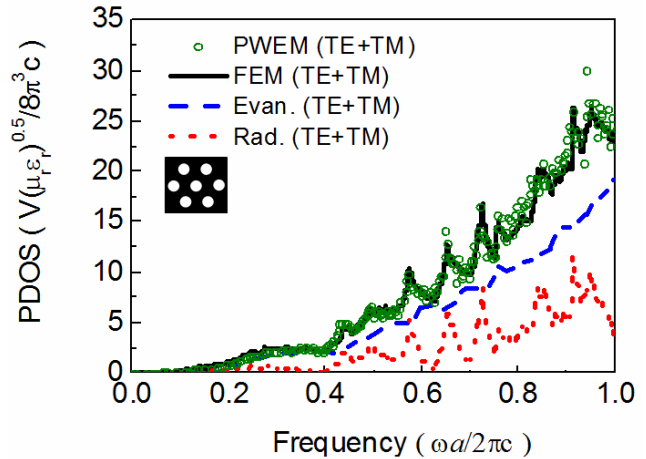}
\\[.3cm]
\caption{\label{fig:epsart} Comparisons of 3D total PDOS calculated by the FEM (black solid line) and the PWEM (green open circles) for a 2D PC with a triangular lattice of air cylinders etched into silicon $(\varepsilon_{r}=11.90)$ at a filling ratio of $67\%$ as used in Ref. \cite{21}. The total PDOS is contributed from both the radiative (red dotted line) and evanescent (blue dashed line) modes. The PBG calculated by the FEM for the off-plane radiative waves ranges from $0.395(2\pi c/a)$ to $0.399(2\pi c/a)$ while that for the in-plane case ranges from $0.382(2\pi c/a)$ to $ 0.400(2\pi c/a)$ \cite{13}.}
\end{figure}

On the other hand, the FEM can be easily adapted to solve problems of great complexity and unusual geometry. The eigenvalues can be accurately and efficiently calculated no matter how complex the geometric structures are, as demonstrated in our previous work \cite{13}. Based on the finite-element analysis of the in-plane dispersion relations of the 2D PCs in the irreducible Brillouin zone, the 3D total PDOS of a 2D PC can be calculated more accurately and efficiently by extending our previous model with the waveguiding theory to consider omnidirectional or off-plane light propagation. In Fig. 2, the PBG calculated by the FEM for the off-plane or omnidirectional  radiative waves ranges from $0.395(2\pi c/a)$ to $0.399(2\pi c/a)$ while that for the in-plane case ranges from $0.382(2\pi c/a)$ to $0.400(2\pi c/a)$ \cite{13}. The PBG diminishes when one considers off-plane or omnidirectional propagation of the radiative modes. However, there is no complete PBG when one also includes the evanescent waves. For demonstration, we further consider two more cases, as illustrated in Figs. 1(a) and 1(b), respectively, including a triangular array with air cylinders etched into a dielectric $(\varepsilon_{r}=12.96)$ at a filling ratio of $75\%$ \cite{19} and a square array with air cylinders etched into a dielectric $(\varepsilon_{r}=12.96)$  at a filling ratio of $72.38\%$ \cite{20}. Both these two specific cases were previously studied and demonstrated to exhibit large omnidirectioinal PBGs. Figures 3(a) and 3(b) show our calculated 3D PDOS for the TE (purple solid lines) and TM (brown solid lines) modes of the triangular array, respectively. Figures 3(c) and 3(d) show those of the square array. The blue dashed and red dotted lines correspond to the PDOS of evanescent and radiative waves. The corresponding PBGs for the radiative waves are determined as (a) $0.310-0.495(2\pi c/a)$ and $0.827-0.834(2\pi c/a)$, (b) $0.403-0.434(2\pi c/a)$, (c) $ 0.410-0.487 (2\pi c/a)$, and (d) $ 0.218-0.259 (2\pi c/a)$ and $ 0.389-0.416 (2\pi c/a)$.

\begin{figure}
\includegraphics[width=0.47\textwidth]{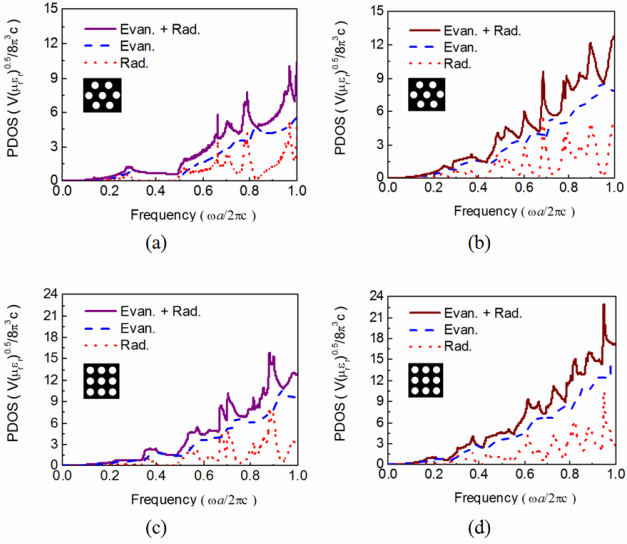}
\\[.3cm]
\caption{\label{fig:epsart} 3D PDOS for (a) TE modes and (b) TM modes of a triangular array with air cylinders etched into a dielectric $(\varepsilon_{r}=12.96)$ at a filling ratio of $75\%$ \cite{19} and for (c) TE modes and (d) TM modes of a square array with air cylinders etched into a dielectric $(\varepsilon_{r}=16.00)$  at a filling ratio of $72.38\%$ \cite{20}. The blue dashed (red dotted) lines correspond to the PDOS of evanescent (radiative) waves. The PBGs calculated by the FEM for the radiative waves are (a) $0.310-0.495(2\pi c/a)$ and $0.827-0.834(2\pi c/a)$, (b) $0.403-0.434(2\pi c/a)$, (c) $ 0.410-0.487 (2\pi c/a)$, and (d) $ 0.218-0.259 (2\pi c/a)$ and $ 0.389-0.416 (2\pi c/a)$.}
\end{figure}

\begin{figure}
\includegraphics[width=0.38\textwidth]{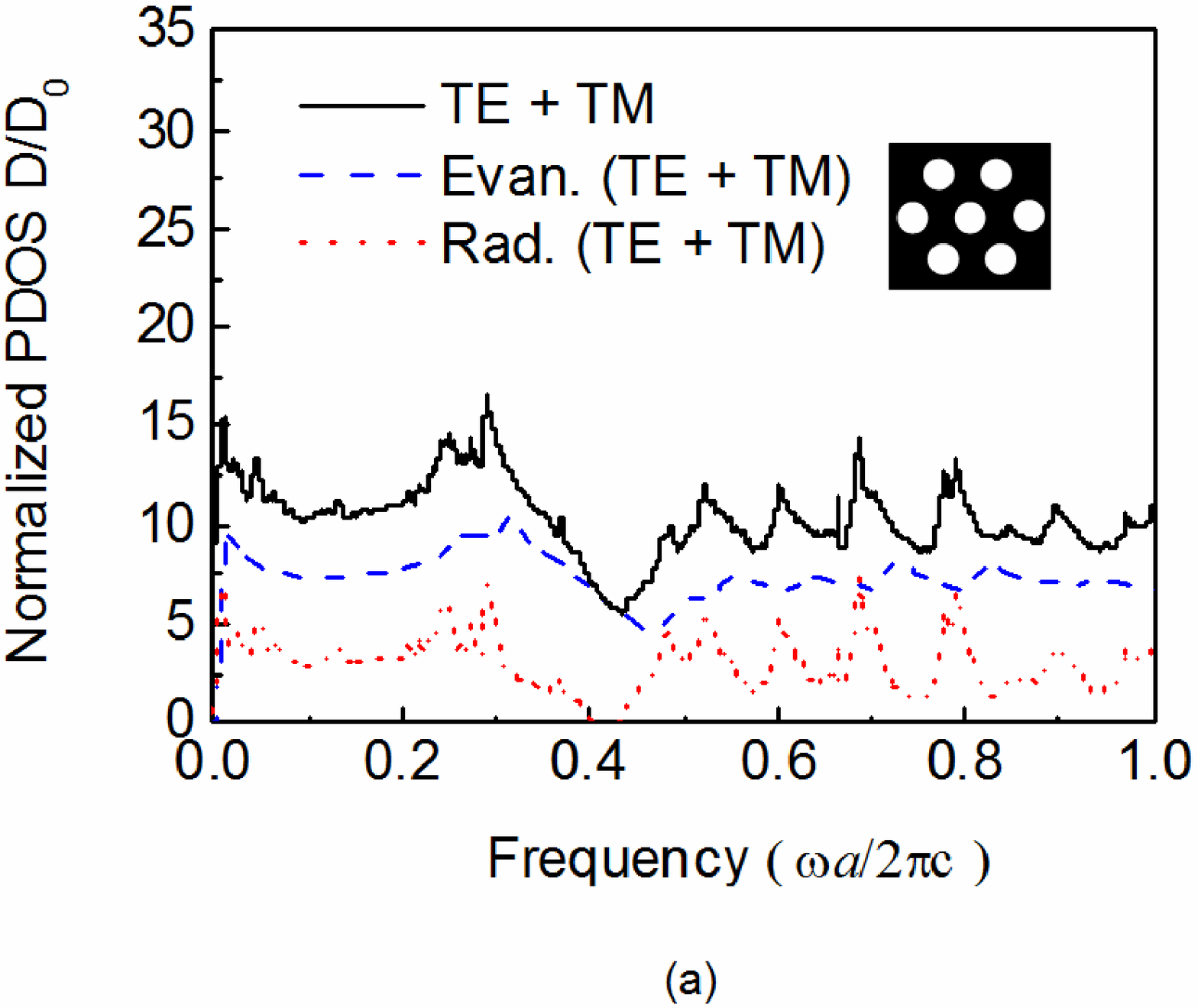}
\\[.3cm]
\includegraphics[width=0.38\textwidth]{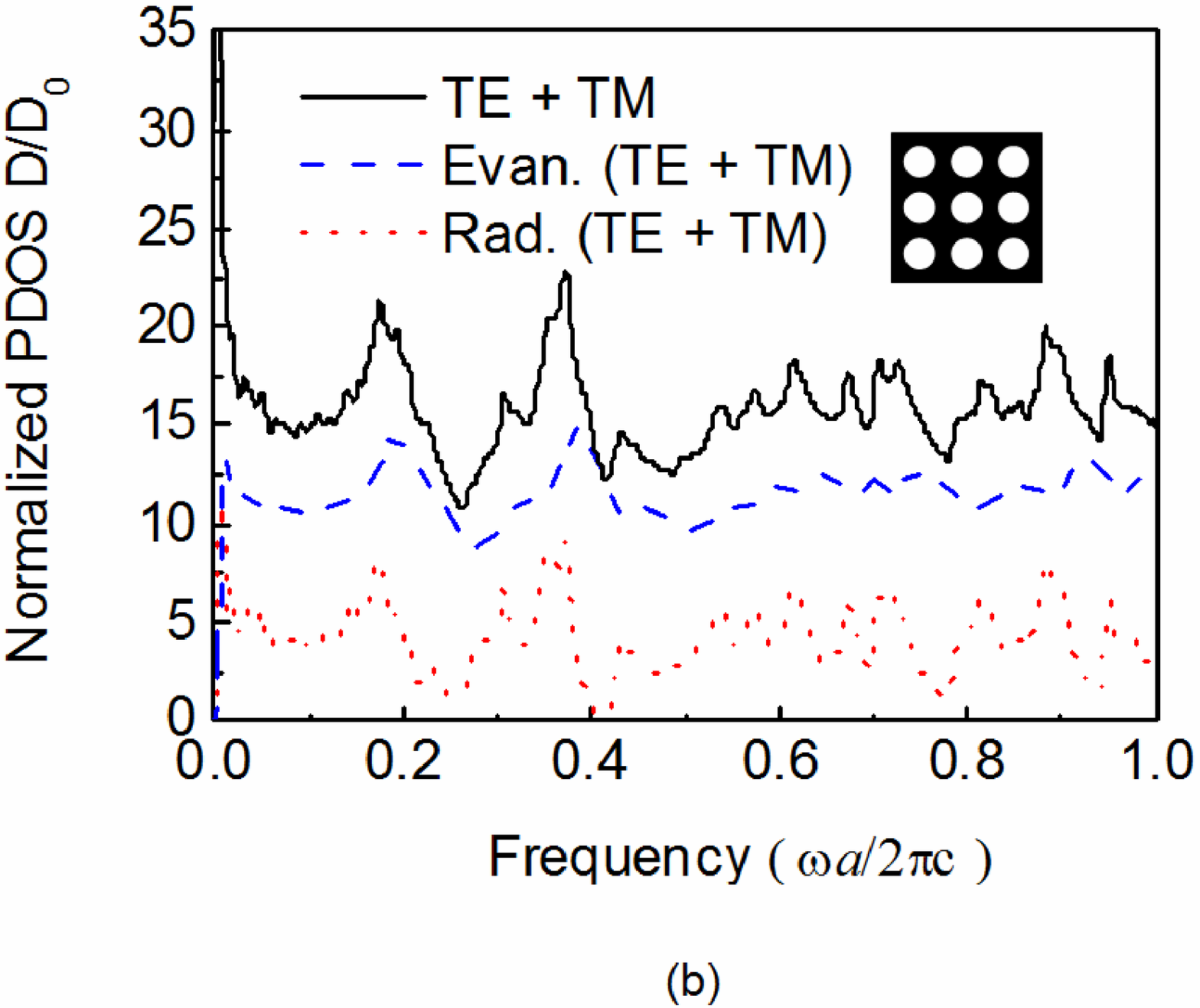}
\caption{\label{fig:epsart} 3D total PDOS (black solid lines) of (a) a triangular array and (b) a square array normalized to that of the vacuum. The blue dashed (red dotted) lines correspond to the normalized PDOS of evanescent (radiative) waves. The PBGs calculated by the FEM for the radiative waves in (a) and (b) range from $0.403 (2\pi c/a)$ to $0.434 (2\pi c/a)$ and from $ 0.410 (2\pi c/a)$ to $0.416 (2\pi c/a)$, respectively. In comparison, those calculated by the PWEM in \cite{19} and \cite{20}range from $0.423 (2\pi c/a)$ to $ 0.437 (2\pi c/a)$ and $0.4045 (2\pi c/a)$ to $ 0.4197 (2\pi c/a)$}
\end{figure}

As one can see, the 3D PDOS of the evanescent waves is larger than that of the radiative waves for both the TE and TM modes. Although the 3D PDOS for the TE and TM modes exhibit similar behavior, the corresponding contributions from the radiative and evanescent parts are quite different. For better understanding spontaneous emission or dipole radiation in a two-dimensional photonic crystal, one may differentiate the PDOS contributed from not only different polarizations, i.e., TE and TM modes, but also different types of waves, i.e., radiative and evanescent waves, by employing our approach.

Figure 4 shows the 3D PDOS normalized to that of the vacuum for the two cases. The blue dashed and red dotted lines correspond to the normalized PDOS of evanescent and radiative waves of both the TE and TM modes. The PBGs calculated by the FEM for the radiative waves in (a) and (b) range from  $ 0.403 (2\pi c/a)$ to $0.434 (2\pi c/a)$ and from $0.410 (2\pi c/a)$ to $0.416 (2\pi c/a)$, respectively. In comparison, those calculated by the PWEM in \cite{19} and \cite{20} range from $0.423 (2\pi c/a)$ to $0.437 (2\pi c/a)$ and $0.4045 (2\pi c/a)$ to $0.4197 (2\pi c/a)$, respectively.  Although the 3D PDOS of the radiative waves for both cases exhibit a PBG, the "complete band gaps" predicted by previous work have been closed by including the contribution of the evanescent modes. Therefore, there is no complete PBG for omnidirectional light propagation in a 2D PC if one considers both radiative and evanescent waves.

\section{\label{sec:level1}CONCLUSIONS}

In summary, omnidirectional light propagation in 2D photonic crystals has been investigated. The polarization characteristics including both the TE and TM modes was considered in our simulation model by extending the formerly developed 2D finite element analysis. The contributions to the 3D total PDOS from the radiative and evanescent waves of different polarizations can be determined separately. We have carefully validated our extended model by benchmarking the results against those calculated by the well-known PWEM, resulting in good agreement. It has been demonstrated that the ¡°complete PBGs¡± shown by previous work considering only the radiative modes will be closed by including the contributions of the evanescent modes. Therefore, a ¡°complete PBG¡± does not exhibit in 2D PCs retaining translational symmetry in the longitudinal direction, if one considers both radiative and evanescent modes. With our approach, an omnidirectional PDOS of 2D PCs can be determined accurately and efficiently. These results are of general importance and relevance to the spontaneous emission by an atom or to dipole radiation in two-dimensional periodic structures. In addition, it may serve as an efficient approach to identifying the existence of a complete PBG in a 2D PC instead of using time-consuming 3D BS calculations.

\begin{acknowledgments}

The authors would like to thank the late Prof. B. Y. Gu at the Institute of Physics, CAS and
Prof. C. T. Chan at the Department of Physics, HKUST for the helpful comments, discussions,
and encouragement. This work was partially supported by Guangdong Industry Polytechnic, P.
R. China, under Grant No. RC201402, the research fund of Hanyang University (HY-
201400000002393), National Research Foundation of Korea (201500000002559), and
the Alexander von Humboldt Foundation of Germany.

\end{acknowledgments}


\begin{thebibliography}{00}

\bibitem{1} J. D. Joannopoulos, S. G. Johnson, J. N. Winn, and R. D. Meade, {\em  Photonic Crystals: Molding the Flow of Light\/}
(Princeton University Press, 2008).
\bibitem{2} J. D. Joannopoulos, P. R. Villeneuve, and S. Fan,  Nature (London) {\bf 386}, 143  (1997).
\bibitem{3} E. M. Purcell, Phys. Rev. {\bf 69}, 681 (1946).
\bibitem{4} D. Kleppner, Phys. Rev. Lett. {\bf 47}, 233 (1981).
\bibitem{5} A. O. Barut and J. P. Dowling, Phys. Rev. A {\bf 36}, 649 (1987).
\bibitem{6} H. Rigneault and S. Monneret, Phys. Rev. A {\bf 54}, 2356 (1996).
\bibitem{7} J. P. Dowling and C. M. Bowden, Phys. Rev. A {\bf 46}, 612 (1992).
\bibitem{8} T. Suzuki and P. K. L. Yu, Opt. Soc. Am. B {\bf 12}, 570 (1995).
\bibitem{9} A. Kamli, M. Babiker, A. Al-Hajry, and N. Enfati, Phys. Rev. A {\bf 55}, 1454 (1997).
\bibitem{10}A. S. S{\'a}nchez and P. Halevi, Phys. Rev. E  {\bf 72}, 056609 (2005).
\bibitem{11}P. Halevi and A. S. S{\'a}nchez, Opt. Commun. {\bf 251}, 109 (2005).
\bibitem{12}M. C. Lin and R. F. Jao, Phys. Rev. E {\bf 74}, 046613 (2006).
\bibitem{13}M. C. Lin and R. F. Jao, Opt. Express  {\bf 15}, 207 (2007).
\bibitem{14}I. A. Sukhoivanov, I. V. Guryev, J. A. Andrade Lucio, E. Alvarado Mendez, M. Trejo-Duran, M. Torres-Cisneros, Microelectronics Journal  {\bf 39}, 685 (2008).
\bibitem{15}Y. C. Tsai, C. F. Lin, and J. W. Chang, Optical Review. {\bf 16}, 347 (2009).
\bibitem{16}Q. Wang, S. Stobbe, and P. Lodahl, Phys. Rev. Lett. {\bf 107}, 167404 (2011).
\bibitem{17}S. R. Huisman, G. Ctistis, S. Stobbe, A. P. Mosk, J. L. Herek, A. Lagendijk, P. Lodahl, W. L. Vos, and P. W. H. Pinkse, Phys. Rev.  B {\bf 86}, 155154 (2012).
\bibitem{18}E. Yeganegi, A. Lagendijk, A. P. Mosk, and W. L. Vos, Phys. Rev. B {\bf 89}, 045123 (2014).
\bibitem{19}Z. Y. Li and Y. Xia, Phys. Rev. B {\bf 64}, 153108 (2001).
\bibitem{20}T. Haas and A. Hesse, T. Doll, Phys. Rev. B {\bf 73}, 045130 (2006).
\bibitem{21}K. Busch and S. John, Phys. Rev. E {\bf 58}3896 (1998).
\bibitem{22}D. P. Fussell, R. C. McPhedran, C. Martijn de Sterke, and A. A. Asatryan, Phys. Rev. E {\bf 67}, 045601(R) (2003).
\bibitem{23}M. M. Sigalas, R. Biswas, K. M. Ho, and C. M. Soukoulis, Phys. Rev. B {\bf 58}, 6791 (1998).
\bibitem{24}S. Foteinopoulo, A. Rosenberg, M. M. Sigalas, and C. M. Soukoulis, J. Appl. Phys. {\bf 89}, 824 (2001).
\bibitem{25}A. Rosenberg, R. J. Tonucci, and E. L. Shirley, J. Appl. Phys. {\bf 82}, 6354 (1997).
\bibitem{26}R. D. Meade , K. D. Brommer, A. M. Rappe, and J. D. Joannopoulos, Appl. Phys. Lett. {\bf 61}), 495 (1992).
\bibitem{27}H. S. S\"{o}z\"{u}er, J. W. Haus and R. Inguva, Phys. Rev. B {\bf 45}, 13962 (1992).

\end{thebibliography}
\end{document}